\documentclass[aps,prl,preprint,groupedaddress,amsfonts,amssymb,amsmath,showpacs]{revtex4}

\usepackage{slashed}

\begin{document}

\title{Possible cosmological implications in electrodynamics due to 
       variations of the fine structure constant}

\author{J.~L. Mart\'{\i}nez-Ledesma$^{1,2,}$}
\email[Email address: ]{albatros@eros.pquim.unam.mx}
\author{S. Mendoza$^{3,}$}
\email[Email address: ]{sergio@astroscu.unam.mx}
\affiliation{$^1$Facultad de Qu\'{\i}mica, Departamento de F\'{\i}sica y
                 Qu\'{\i}mica Te\'orica, Universidad Nacional Aut\'onoma
                 de M\'exico, Circuito Interior, Distrito Federal 04510, 
		 M\'exico\\
             $^2$Universidad de la Ciudad de M\'exico, San 
                 Lorenzo 290, Colonia del Valle, Distrito Federal
		 03100, M\'exico\\
             $^3$Instituto de Astronom\'{\i}a, Universidad Nacional
                 Aut\'onoma de M\'exico, AP 70-264, Distrito Federal 04510,
	         M\'exico
            }

\date{\today}

\begin{abstract}
  Astronomical observations are suggesting that the fine
structure constant varies cosmologically.  We present an analysis on the
consequences that these variations might induce on the electromagnetic
field as a whole.  We show that under these circumstances the
electrodynamics in vacuum could be described by two fields, the ``standard''
Maxwell's field and a new scalar field.  We provide a generalised Lorentz
force which can be used to test our results experimentally. 
%
\vskip+1cm
\centerline{\large Resumen}
  Observaciones astron\'omicas sugieren que la constante de estructura
fina presenta variaciones cosmol\'ogicas.  En este art\'{\i}culo hacemos
un an\'alisis sobre las consecuencias que estas variaciones posiblemente 
inducen en el campo electromagn\'etico.  Mostramos que bajo estas circunstancias
la electrodin\'amica del vac\'{\i}o puede ser descrita por dos campos,
el campo ``est\'andar'' de Maxwell y un nuevo campo escalar.  Adem\'as,
proponemos una fuerza de Lorentz generalizada que puede  utilizarse para
confirmar nuestros resultados de manera experimental.
\end{abstract}


\maketitle

\section{Introduction}

  Since the first half of the 20th century different researchers
\cite{dirac38,teller48,jordan59} began to put forward the idea that
the fine structure constant \( \alpha \) could present cosmological
variations.  Today, recent observations of quasars have suggested
\cite{webb99,webb01,murphy01} that the fine structure constant \( \alpha
\equiv e^2 / \hbar c \) might present variations with respect to cosmic
time.  Here, \( e \) represent the electron charge, \( c \) the speed
of light and \( \hbar \) Planck's constant. These observations imply
that the fluctuations \( \Delta \alpha / \alpha \) of this fundamental
constant are given by

\begin{equation}
  \frac{ \Delta \alpha }{ \alpha } = -0.72 \pm 0.18 \times 10^{-5},
\label{eq.1}
\end{equation}

\noindent in the interval of redshifts \( z \) given by \( 0.5 < z <
3.5 \).

  Because the constants \( \hbar,\ e \text { and } c \) that define \(
\alpha \) might vary in different ways \cite{davies02} so as to give
the value given by eq.\eqref{eq.1}, one may assume that the electromagnetic
fields and the electric charges are coupled in different forms that
depend on cosmic time. Since the electric charge could have variations
of cosmological origin, possibly the continuity equation no longer holds
and/or part of the electric charge is not generating electromagnetic
field, or contrary, it generates an extra electromagnetic field.
In this letter we explore these possibilities and some of its immediate
consequences on space--time.

  Previous research has been conducted on this topic.  Most notably
the work by Bekenstein \cite{bekenstein82} and Chodos \& Detweiler
\cite{chodos80} had given in the past theoretical clues as to why
the fine structure constant might vary in time or position as the
universe expands.  Bekenstein developed a complete
analysis using the principle of least action.  Chodos \& Detweiler
analysed \( \alpha \) variations using a five dimensional
(4+1) space--time based on ideas first proposed by Kaluza and Klein.

  It is well known that one can decompose a vector field as the sum of one 
solenoidal component plus a non--rotational one (cf. Helmholtz
decomposition theorem).  A generalisation in terms of differential forms is
given by the Hodge decomposition theorem for Riemannian manifolds.  Also,
an \( n \)--dimensional  manifold can be foliated with submanifolds of
smaller dimensions.  For the electromagnetic case that we study in this
letter, it is possible to foliate the space--time (with a 3+1 Lorentzian
metric) with 2--dimensional and 0--dimensional manifolds that ``emerge''
from vector fields which represent the electric charge--current densities.
It is then natural to use the formalism of differential forms in order to
obtain a more general study of the problem through a Hodge--like
decomposition of the differential form that represents the electromagnetic 
charge--current distributions.

\section{Electrodynamics}

  Let us take the 1--form \( \boldsymbol{J}_\text{std} = \rho_\text{std}
\, \boldsymbol{\mathrm{d}} x^0 + \left( j^{\text{(std)}}_k / c \right)
\boldsymbol{\mathrm{d}} x^k \)  representing the charge--current
in the usual sense \cite{MTW} with \( k = 1,\, 2,\, 3 \). Here the signature
of the metric is given by (\(-,\,+,\,+,\,+\)),  \(
\rho_\text{std} \) is the charge density, \( j^{\text{(std)}}_k \)
are the components of the current density, \( \boldsymbol{\mathrm{d}}
x^\mu \) is a basis for the cotangent space with coordinates (\( x^0
\! = \! ct,\, x^1,\, x^2,\, x^3 \)) and greek indices have values \(
0,\, 1,\, 2,\, 3 \). \( \boldsymbol{J}_\text{std} \) satisfies
Maxwell's equations

\begin{equation}
  \boldsymbol{\mathrm{d}} \boldsymbol{ F } = 0, \qquad \boldsymbol{\delta} \boldsymbol{  F } =
    4 \pi \boldsymbol{J}_\text{std},
\label{eq.2}
\end{equation}

\noindent which imply naturally the continuity equation

\begin{equation}
  \boldsymbol{\delta} \boldsymbol{J}_\text{std} = 0.
\label{eq.3}
\end{equation}

\noindent In the previous equations \( \boldsymbol{F} \) is a 2--form that
builds up  the standard electromagnetic field and is given \cite{MTW}
by \( \boldsymbol{ F } \equiv E_1 \boldsymbol{\mathrm{d}} x^1 \wedge
\boldsymbol{\mathrm{d}} x^0 + E_2 \boldsymbol{\mathrm{d}} x^2 \wedge
\boldsymbol{\mathrm{d}} x^0 + \ldots + B_3 \boldsymbol{\mathrm{d}}x^1
\wedge \boldsymbol{\mathrm{d}}x^2 \).  \( \boldsymbol{E} \text{
and } \boldsymbol{B} \) represent the electric and magnetic
components of the electromagnetic field. \( \boldsymbol{\delta}
\equiv ^\star \!  \boldsymbol{\mathrm{d}} ^\star \)  is the
\mbox{co--differential} operator and \( ^\star \) is the Hodge star
operator \cite{misner57,MTW,nakahara90,derham}.

  In a universe with varying \( \alpha \), the continuity equation
is not necessarily valid. This can be interpreted as if a universal
``total charge--current'' 1--form \( \boldsymbol{J}_e \) given by

\begin{equation}
  \boldsymbol{J}_e \equiv \boldsymbol{J}_\text{std} +
    \boldsymbol{J}_\text{n} + \boldsymbol{J}_\text{h},
\label{eq.4}
\end{equation}

\noindent is associated to the ``global electrodynamics'' of the
universe at all cosmological times.  In eq.\eqref{eq.4}, the 1--form \(
\boldsymbol{J}_e \) is such that it accepts a Hodge--like decomposition
(cf. Hodge decomposition theorem in \cite{nakahara90,derham}).  With this
assumption, the differential 1--forms \( \boldsymbol{J}_\text{std} \),
\( \boldsymbol{J}_\text{n} \) and \( \boldsymbol{J}_\text{h} \)  are
coexact, exact and harmonic 1--forms respectively.

  Eq.\eqref{eq.4} is a natural generalisation of the result expressed by
eq.\eqref{eq.1}.  Indeed, eq.\eqref{eq.1} means that \( \alpha \approx
\left( 1 - 0.72 \times 10^{-5} \right) \alpha_\text{today} \).  If the
total charge--current \( \boldsymbol{J}_e \) obeys a similar relation,
that is

\begin{equation}
  \boldsymbol{J}_e = ( 1 + \eta ) \boldsymbol{J}_\text{std},
\label{eq.5}
\end{equation}

\noindent where \( \eta \) is a scalar 0--form, then it follows
that \( \boldsymbol{J}_\text{n} + \boldsymbol{J}_\text{h} = \eta
\boldsymbol{J}_\text{std} \).  To simplify things it is possible to
assume that \( \eta  \) can be decomposed in two additive terms, \(
\eta_\text{n} \) and \( \eta_\text{h} \) such that \( \eta = \eta_\text{n}
+ \eta_\text{h} \). These terms satisfy

\begin{equation}
  \boldsymbol{J}_\text{n} = \eta_\text{n} \boldsymbol{J}_\text{std},
    \qquad \text{ and } \qquad \boldsymbol{J}_\text{h} = \eta_\text{h}
    \boldsymbol{J}_\text{std}.
\label{eq.6}
\end{equation}

  From the previous considerations it follows that the \mbox{1--form} \(
\boldsymbol{J}_e \) does not satisfy a continuity--like equation when \(
\eta_n \neq 0 \).

\section{Mathematical relations between fields}

  In order to analyse the electrodynamics imposed by the conditions of
the previous section, let us multiply eq.\eqref{eq.4} by \( 4 \pi \)
and substitute eq.\eqref{eq.2} and eq.\eqref{eq.6} on this to obtain

\begin{equation}
  4 \pi \boldsymbol{J}_e = \boldsymbol{\delta} \boldsymbol{F}
    + \boldsymbol{\mathrm{d}} M + 4 \pi \eta_\text{h}
    \boldsymbol{J}_\text{std},
\label{eq.7}
\end{equation}

\noindent in which the scalar 0--form \( M \) is such that

\begin{equation}
  \boldsymbol{\mathrm{d}} M =  4 \pi \eta_\text{n}
    \boldsymbol{J}_\text{std}, \qquad \text{ and } \qquad
    \boldsymbol{\delta} M = 0.
\label{eq.8}
\end{equation}

  Note that eq.\eqref{eq.7} reduces to the standard Maxwell's equations
when there is no cosmological variation of \( \boldsymbol{J}_e \). That is,
when \( \eta_\text{n} = \eta_\text{h} = 0 \) and so \( \boldsymbol{J}_e =
\boldsymbol{J}_\text{std} \).  In the general case, when this condition
is not valid, the electromagnetic field is such that it is represented by
two mathematical objects, the Maxwell 2--form \( \boldsymbol{F} \) and
the 0--form \( M \).  \( \boldsymbol{F} \) satisfies Maxwell's equations,
eq.\eqref{eq.2}, and \( M \) satisfies a set of Maxwell's-like equations
given by eq.\eqref{eq.8}.  In other words, the cosmic time variations
of \( \boldsymbol{J}_e \) imply that the electrodynamics of space--time
are given by two fields.  One field turns out to be the standard
Maxwell 2--form \( \boldsymbol{F} \).  The other is a scalar field \(
M \) introduced by the cosmological variations of \( \boldsymbol{J}_e \).

  According to eq.\eqref{eq.7}, the 0--form \( M \) satisfies the
following ``Poisson's'' equation

\begin{equation}
  \Delta M \equiv \left( \boldsymbol{\delta} + d \right)^2 M =
    ^\star \! \! \left\{ \boldsymbol{\mathrm{d}} \eta_\text{n} \wedge
    \boldsymbol{\mathrm{d}} ^\star \! \boldsymbol{F} \right\}.
\label{eq.9}
\end{equation}

\noindent In other words, the scalar field \( M \) is produced by the
changes in the 2--form field \( \boldsymbol{F} \) and the scalar \(
\eta_\text{n} \).

  We can also give an expression for Dirac's equation.  From
eq.\eqref{eq.7}, using again a Hodge--like decomposition,  it follows
that  we can introduce a 1--form \( \boldsymbol{A} \) that represents
the electromagnetic potential given by

\begin{equation}
  \boldsymbol{A} = \boldsymbol{A}_\text{std} + \boldsymbol{A}_M +
    \boldsymbol{A}_\text{h}, 
\label{eq.9.1}
\end{equation}

\noindent  where \( \boldsymbol{A}_\text{std},\,  \boldsymbol{A}_M
\text{ and } \boldsymbol{A}_\text{h} \) are co--exact, exact and harmonic
1--forms respectively. In eq.\eqref{eq.9.1} we have added the 1--form \(
\boldsymbol{A}_\text{h} \) for mathematical completeness, despite the
fact that it is usually discarded in standard physics.  With this, and
because \( e = ( 1 + \eta ) e_\text{std} \), where \( e_\text{std} \) is
the standard charge of an electron, then Dirac's equation takes the form

\begin{equation}
   \left( i \, \boldsymbol{\slashed{\mathrm{d}}} - \frac{ \alpha }{
    \left( 1 + \eta \right) e_\text{std} } \boldsymbol{\slashed{A}}
    \right) \Psi = \frac{ m c }{ \hbar } \boldsymbol{1} \Psi.
\label{eq.9.2}
\end{equation}

\noindent Here  \( i^2 = -1 \), \( \boldsymbol{\slashed{\mathrm{d}}}
= \gamma^\mu \partial_\mu \), \( m \) is the electron's rest mass, \(
\boldsymbol{\slashed{A}} = \gamma^\mu A_\mu \), \( \Psi \) is Dirac's
spinor and  \( \boldsymbol{1} \) is the identity element of the algebra
generated by Dirac's matrices \( \gamma^\mu \) that satisfy the
following equation

\begin{displaymath}
  \gamma^\mu \gamma^\nu + \gamma^\nu \gamma^\mu = 2 g^{\mu\nu} \,
    \boldsymbol{1},
\label{eq.9.3}
\end{displaymath}

\noindent where \( g^{\mu\nu} \) are the metric components assigned to
space--time.

\section{Discussion}

 The previous analysis was made under the assumption that the variations
of the 1--form \( \boldsymbol{J}_e \) are time dependent.  However, all
the presentation is still valid if the variations are not only functions that
depend on time, but also functions that could vary on space.  That is,
the variations can equally occur on space and/or time and the coupling
of the two fields \( \boldsymbol{F} \) and \( M \) will still occur in
the same form.  More generally, the results obtained in the previous
section are also valid if space--time variations on the ``fundamental''
constants \( \hbar,\, e \text{ and } c \), or even \cite{ivanchik02} 
\( m \)  occur.

  It is intriguing that our daily experiments do not show any
evidence of the physical properties that the field \( M \)
might induce on space--time.   However, there has been a report
\citep{monstein02,vlaenderen03} in which such a field produces
longitudinal electrodynamic waves.  One can also think that the reason
for a non--observable field \( M \) happens because it vanishes at
our present epoch.  This is the same as saying that we live in a very
peculiar place or time in the universe, something that is difficult
to believe.  On the other hand, one can think that we have constructed
our standard Maxwell electrodynamics in such a way that the properties
of the field \( M \) do not affect any of our experiments.  This is also
difficult to believe.  Another  possible way in which the field \( M \)
might had been missed by our experiments is if its strength is tiny.
For example, since eq.\eqref{eq.1} suggests that \( \eta \) is a small
quantity, then it follows that the field \( M \) is weak.  Indeed,
when \( \eta = 0 \) then \( \eta_\text{h} = - \eta_\text{n} \). This
result together with eq.\eqref{eq.6} and combined with the properties
of \( \boldsymbol{J}_\text{n} \text{ and } \boldsymbol{J}_\text{h} \)
imply that \( \eta_\text{h} = \eta_\text{n} = 0 \).  Thus, the trivial
solution of eq.\eqref{eq.8} occurs when \( \eta = 0 \) and  gives
\( M = 0 \) because \( M \) is not harmonic.  When \( \eta \) is a
small quantity, one has to proceed slightly different.  The Lorentz
force can be naturally  generalised as \( \mathrm{d} \boldsymbol{P}
/ \mathrm{d} \tau = {}^*\!\boldsymbol{F} \cdot {}^*\!\boldsymbol{J}_e +
M \boldsymbol{J}_e = \left( 1 + \eta \right) \left( {}^*\!\boldsymbol{F}
\cdot {}^* \! \boldsymbol{J}_\text{std} + M \boldsymbol{J}_\text{std}
\right) \), where \( \tau \) is the proper time and \( \boldsymbol{P}
\) is the 1--form momentum. So, if \( \eta \) is small and \( M \) is
not negligible then we would had already observed the properties of the
field \( M \) in our laboratories. However, this Lorentz force can be
used in experiments to test the validity of our reasoning.

  On the other hand, when \( \eta_\text{n} = 0 \), then \(
\boldsymbol{M} = 0 \),  and the Lorentz force is given by 

\begin{equation}
  \frac{ \mathrm{d} \boldsymbol{P} }{  \mathrm{d} \tau } = 
    \left( 1 + \eta_\text{h}\right) {}^*\! \boldsymbol{F} \cdot 
    {}^* \! \boldsymbol{J}_\text{std}.
\label{eq.10}
\end{equation}

\noindent This means that the standard Lorentz force is changed by a factor
\( ( 1 + \eta_\text{h} ) \) because the variations of \( \eta_\text{h}
\) produce deviations in the intensities of the electromagnetic
interactions.

  However, eq.\eqref{eq.10} can be written as

\begin{equation}
  \frac{ 1 }{ \left( 1 + \eta_\text{h}\right) } \frac{ \mathrm{d}
    \boldsymbol{P} }{  \mathrm{d} \tau } = {}^* \! \boldsymbol{F} \cdot
    {}^* \! \boldsymbol{J}_\text{std}.
\label{eq.11}
\end{equation}

\noindent This equation means that the electromagnetic forces are
producing deviations from the standard dynamics, since \( \eta_\text{h}
\neq 0 \) associates changes on the momentum which are not Newtonian.

  The duality presented in eqs.\eqref{eq.10}-\eqref{eq.11}
is similar to that presented by some researchers
\cite{milgrom83,milgrom01,milgrom02,sanders02} for the gravitational
forces in order to explain the rotation curves of galaxies, and other
astronomical observations.  These theories, the so called Modified
Newtonian Dynamics (MOND) theories, suggest that our standard ideas of
dynamics should be changed.  For the electromagnetic case considered in
the present article, this modification occurs naturally.


\bibliography{charge}
\bibliographystyle{apsrev}

\end{document}